\documentstyle[twoside,fleqn,espcrc2]{article}

\input epsf

\def\myfig#1#2#3{  
\begin{figure}[htb]
\mbox{
  \epsfxsize=70mm \epsfbox{#1}  
}
\vspace{-0.8cm}
\caption{#2}
\label{#3}
\end{figure}
}
 
\newcommand{\figGyration}{
  \myfig{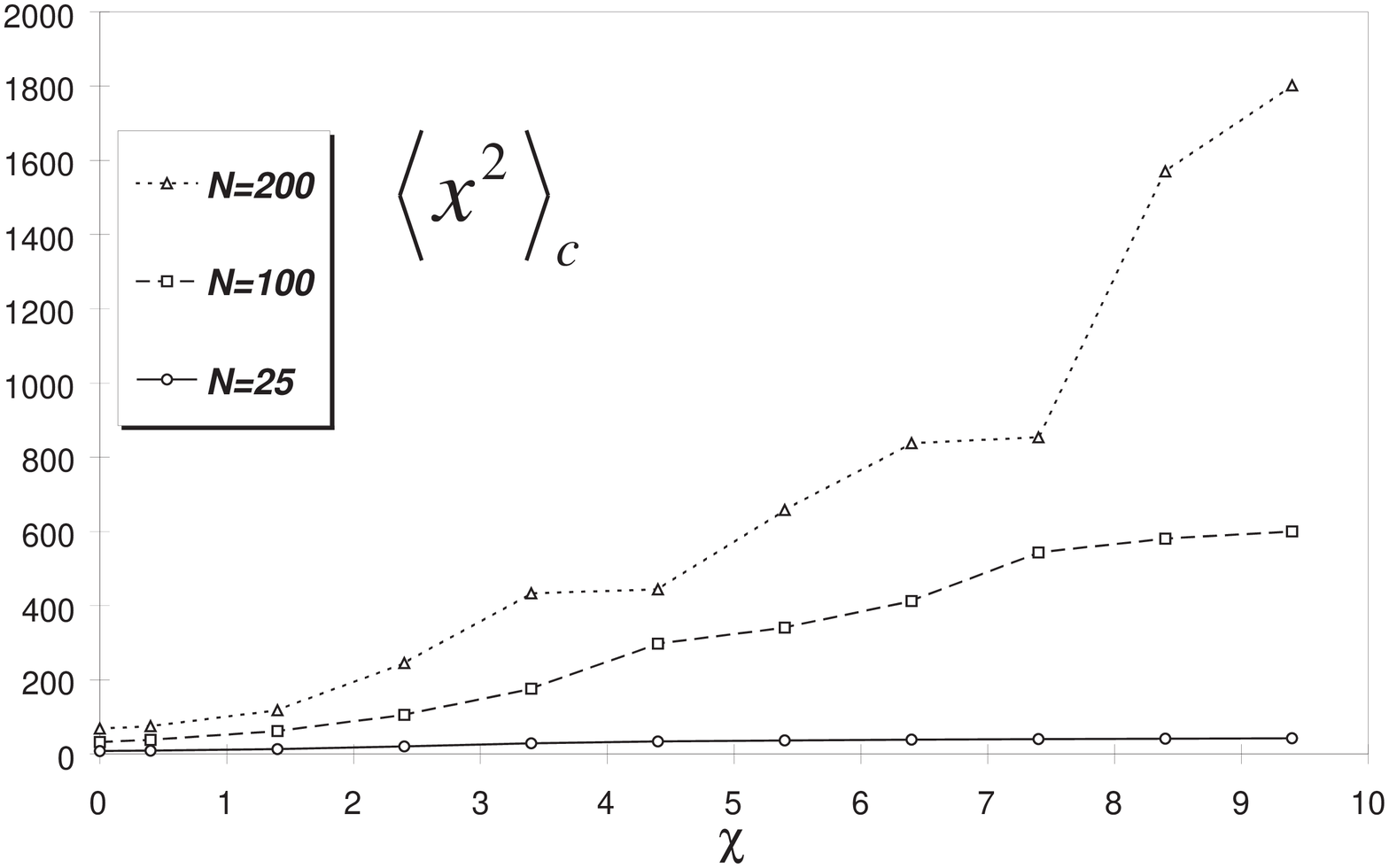}{The gyration radius for several paths}{fig:Gyration}
}
\newcommand{\figHausdorff}{
  \myfig{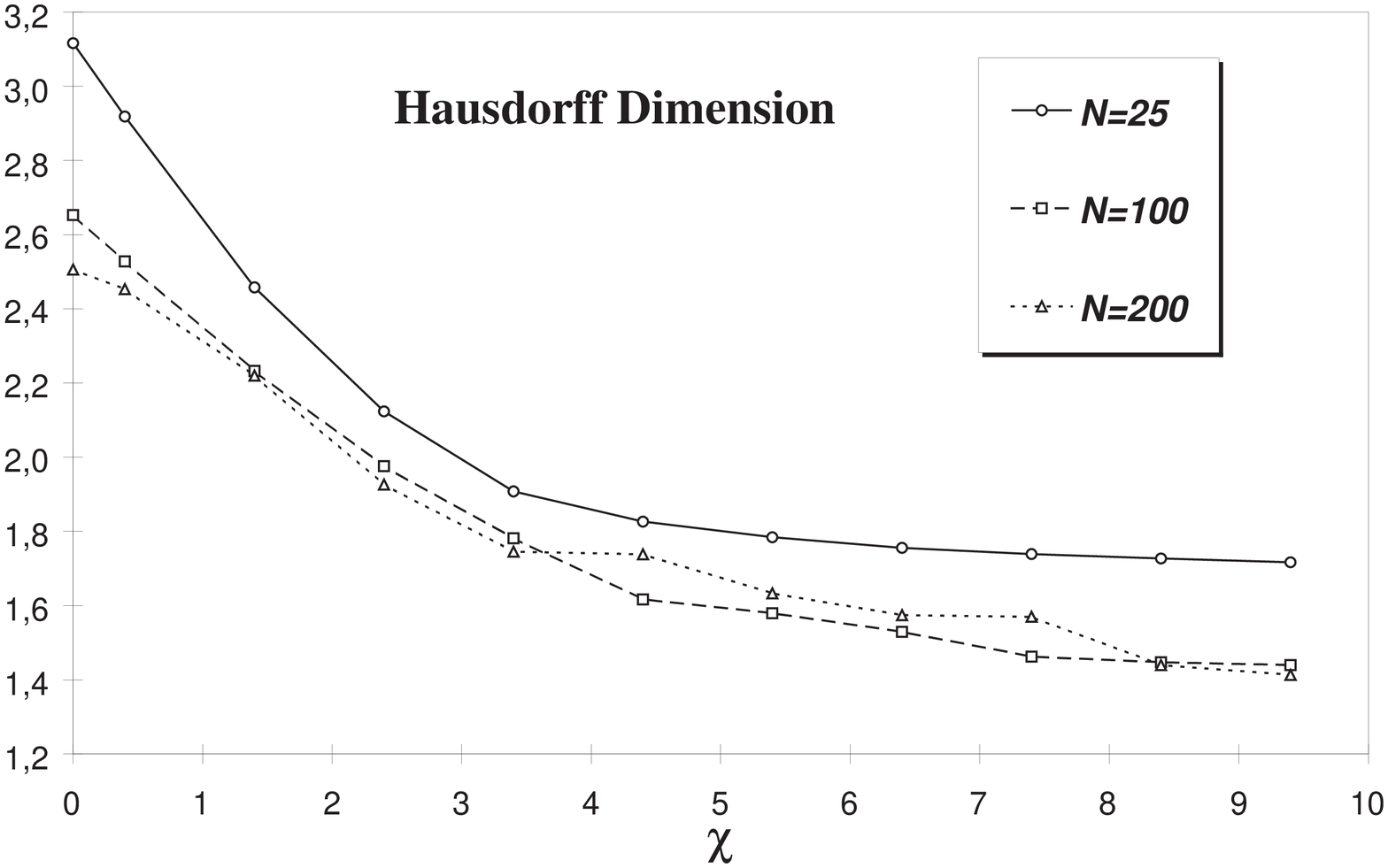}
   {The Hausdorff dimension for several paths}{fig:Hausdorff}
}

\newcommand{\figCv}{  
  \myfig{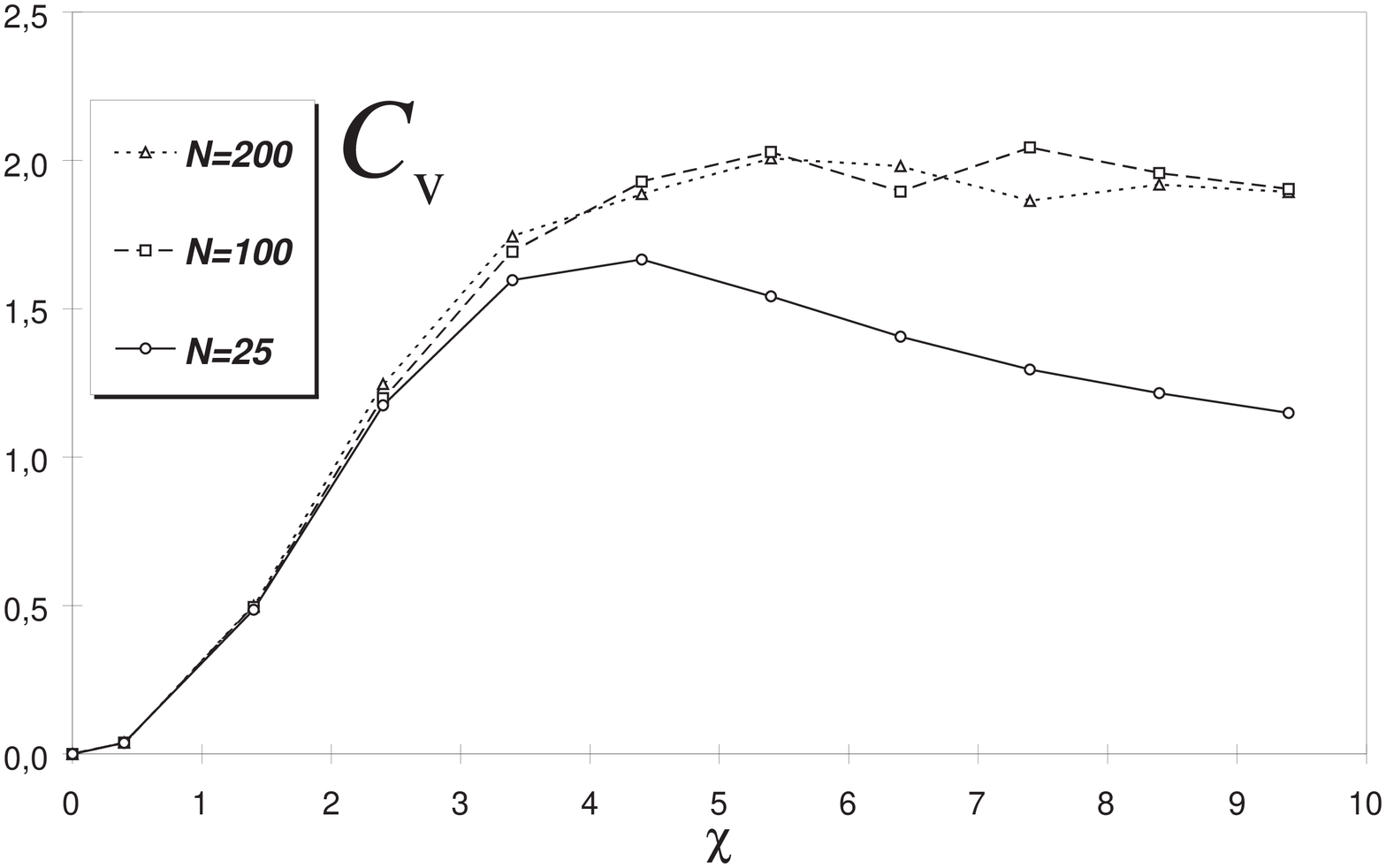}{The specific heat for several paths}{fig:Cv}
}

\newcommand{\figCurvature}{  
  \myfig{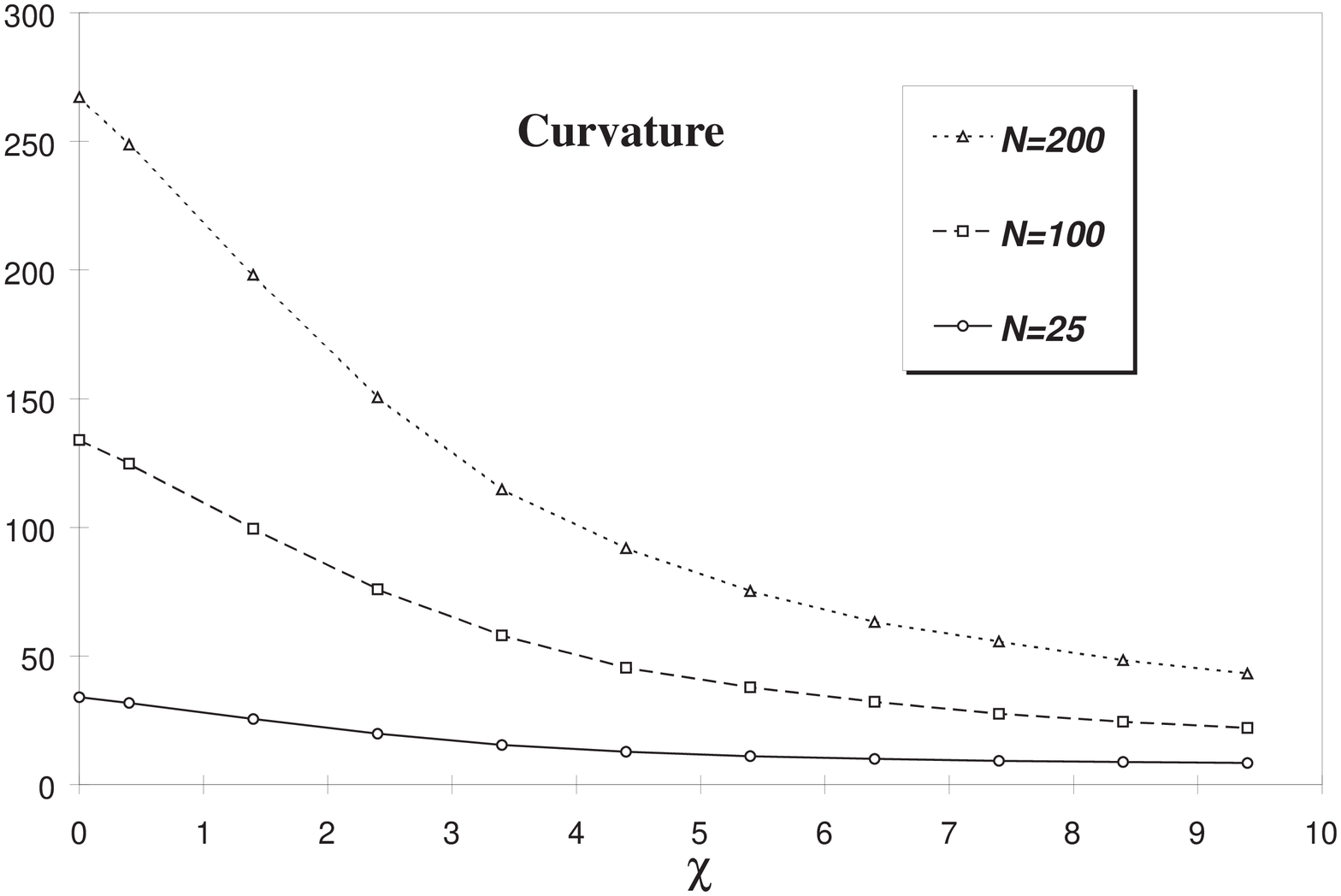}{The curvature for several paths}{fig:Curvature}
}

\def\figSnapShot{  
\begin{figure*}[ht]
\centerline{
\mbox{
  \epsfysize=11cm \epsfbox{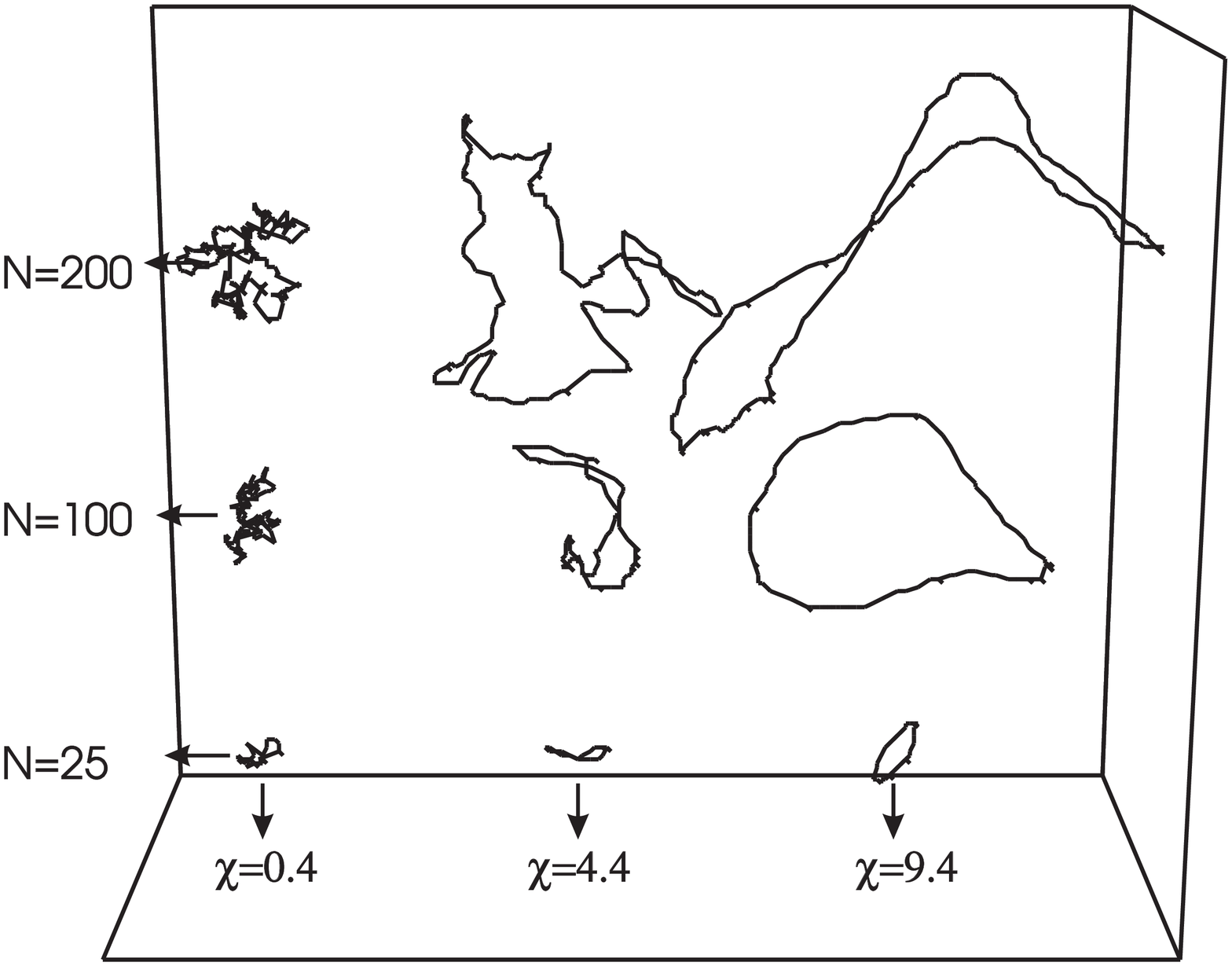}    
}}
\vspace{-0.8cm}
\caption{Snapshots for several paths. Notice how the ``size'' of the
paths (the gyration radius) grows with~$N$ and how the curvature
decreases as $\chi$ grows.}
\label{fig:SnapShot}
\end{figure*}
}
  
\title{ \hfill {\normalsize  UAB--FT/396} \\
        \hfill {\normalsize hep--lat/9607040}\\
Random paths with curvature}

\author{M. Baig
        \thanks{e-mail:{\tt baig@ifae.es},
                URL:{\tt http://www.ifae.es/$\sim$baig}}
        and 
        J. Clua 
        \thanks{e-mail:{\tt clua@ifae.es},
                URL:{\tt http://www.ifae.es/$\sim$clua} }
        \\[15pt]
        Institut de F\'{\i}sica d'Altes Energies (IFAE). 
        Facultat de Ci\`encies. Edifici Cn\\
        Universitat Aut\`onoma de Barcelona.
        08193 Bellaterra (Barcelona) Spain
}

\begin{document}

\begin{abstract}
We present some results coming from a Monte Carlo simulation of a set
of random paths with a curvature dependent action. This model can be 
considered as a toy model of the theory of random surfaces.
The transition from free to rigid random paths has been analyzed and
the similitude with the crumpling transition have been pointed out.
\end{abstract}


\maketitle

\section{Introduction}
Many  problems that arise in physics can be related to the properties
of random walks and random surfaces: 
  \begin{itemize} 
  \item The Feynman formulation of Quantum Mechanics.
  \item The 3-d Ising model. 
  \item The 2-d quantum gravity. 
  \item The behavior of interfaces in mixtures.
  \item The crystal growth.
  \item The behavior of polymers.
  \end{itemize}

\section{The Physics of Random Surfaces}
The study of random surfaces as a generalization of Brownian motion, 
led to a renewed interest after the work of A. N. Polyakov\cite{Pol81}.
In 1984, A. Billore, D.J. Gross and E. Marinari\cite{BGM84} applied 
Monte Carlo techniques to the numerical study of free random surfaces 
defined as a set of a fixed number of triangles embedded in a 
continuum space. This method, in practice, is a simulation of a 
microcanonical ensemble of closed random surfaces with an action 
proportional to the area. However, they showed that the generated 
surfaces had an anomalous Hausdorff dimension.

To overcome this problem, several authors investigated\cite{PL85} 
the consequences of the addition of an extrinsic curvature term to 
the Nambu-Goto action (i.e. the area action). They suggested that 
this extra term may control the formation of {\em spikes}, 
deformations that are exact zero-modes of the area action
and originates the degeneration of the random surface into branched
polymers.

During last years, different numerical studies on the behavior of an
ensemble of rigid triangulated random surfaces with different
implementations of the extrinsic curvature term have been carried out
in order to determine the nature of the phase transition (the crumpling
transition) that separates the brownian phase and the fixed 
one\cite{BET94}. In addition, several different new approaches, as 
the fluid random surfaces, have been also extensively developed. 

\section{The introduction of Random Paths}

As a toy model, Billoire et al.\cite{BGM84} simulated also an 
ensemble of closed random paths. As a result they concluded that the 
random walks obtained a Hausdorff dimension of 2, in accordance with 
the expectative coming from Brownian studies. Furthermore, R.D. 
Pisarski\cite{Pis86} proposed the addition of a term proportional to 
the curvature also for the theory  of random paths, claiming that 
this theory might be relevant to the polymer physics. 
He noticed the observation of asymptotic freedom and pointed out the
similarities between this theory and a nonlinear $\sigma$-model with
long-range interactions. In addition, F. Alonso and 
D. Espriu\cite{AE87}, in his mean field analysis of random surfaces, 
they included also the analysis of random paths as a simple version 
of the random surface theory.

Historically, random walks have been considerable studied in the 
field of polymer physics\cite{WGB85}. Much less work has been devoted 
to the random paths theories in the context of field theories. 
In 1987 J. Ambj\o rn, B. Durhuus and T. Jonsson\cite{ADJ87} initiated 
an analytic study of random paths with a curvature dependent action. 
First, they considered bosonic paths using two different 
regularizations, namely, random walks on the lattice $Z^d$ and also 
paths consisting of straight line segments in $R^d$, 
i.e. the toy model considered as a simple model from random surfaces.
In a second step they introduced also fermionic paths\cite{ADJ90}.

\section{Our analysis}

We have performed the first numerical analysis of a set of  random paths
with a curvature dependent action\cite{BCJ96}. In addition to the 
intrinsic interest of this study, such a simulation can also be 
considered as a simple case to contrast the numerical work performed 
in the simulation of crystaline random surfaces and, in particular to 
compare with the analysis of the nature of the crumpling transition. 
The main motivation for the numerical study is to determine if there 
is a phase  transition separating the phase of brownian  paths (small 
curvature coupling) and the phase of rigid paths (large coupling). 
This transition would be the analogous of the {\em crumpling 
transition}  observed in the simulation of random surfaces.

\subsection{Lattice action for closed paths in $R^3$}
  
The actual action we have simulated is a lattice transcription of the
action obtained simply replacing the derivatives of the path
$X^m_i$ by finite differences $X^m_i-X^m_i$ (and doing analogously with
second derivatives):
\begin{equation}
 S=\beta\sum_{i=1}^{N} |X_i - X_{i-1}| + 2\chi\sum_{i=1}^{N}|\sin
\frac{\theta_i}{2}|,
\end{equation}
where $N$ is the number of nodes in the path and $\theta_i$ is the angle
between the straight segments that share a node. It is
easy to check that, like in the surface case, this action is invariant
under reparametrizations (which may be thought as a kind of gauge
symmetry). This fact allows to fix the coupling $\beta=1$ obtaining,
actually, an one parameter action.

\subsection{Numerical simulation}

The numerical computation of the partition function has been done 
applying a Monte Carlo simulation using the Metropolis algorithm. 
The data of the simulation are as follows:

\begin{itemize}
  \item Number of points in the paths:\\$N=25, 100, 200$.
  \item Coupling   $\chi$ from 0 to 10,  $\beta=1$ (fixed).
  \item Number of sweeps:  2-4 millions for each coupling.
\end{itemize}
  
The following magnitudes have been measured:

\begin{enumerate}

\item 
As a test of the numerical procedure we have computed first the mean
length of the path. It is easy to deduce that, independently from
the curvature coupling $\chi$,

\begin{equation}
<L>=-\frac{\partial \ln Z}{\partial\beta}=\frac{d(N-1)}{\beta}.
\end{equation}
Results are summarized in the table
$$
\begin{array}{|c|c|c|c|}
\hline
\chi & L=25        & L=100        & L=200         \\ \hline
0.0  & 71.9\pm 0.1 & 296.7\pm 0.2 & 597.3\pm 0.4  \\
1.4  & 71.7\pm 0.1 & 296.5\pm 0.2 & 595.9\pm 0.5  \\
3.4  & 72.3\pm 0.2 & 297.5\pm 0.6 & 600.1\pm 1.0  \\
5.4  & 71.6\pm 0.2 & 297.5\pm 0.8 & 594.8\pm 1.6  \\ \hline
\end{array}
$$

\item
The  mean curvature, which is given in fig.~\ref{fig:Curvature}:
\begin{equation}
<S_c> = -\frac{\partial \ln Z}{\partial \chi}.
\end{equation}

\figCurvature

\item
The heat capacity respect to curvature, defined as
\begin{equation}
C_\chi = \frac{-\chi^2}{N}\frac{\partial^2 \ln Z}{\partial\chi^2},   
\end{equation}
which is shown in fig.~\ref{fig:Cv}.

\figCv

\item
The gyration radius, shown in figure~\ref{fig:Gyration}:
\begin{equation}
<X^2>_c=\frac{1}{d} \left( <X^2>-<X>^2 \right).
\end{equation}

\figGyration

\item
Finally, the Hausdorff dimension of the path defined as 
(see fig.~\ref{fig:Hausdorff}):
\begin{equation}
d_H=\lim_{N\rightarrow \infty} 2 \frac{\ln<X^2>_c}{\ln N}.
\end{equation}

\figHausdorff

\end{enumerate}

\section{Numerical results and conclusions}

There appear two different regimes:
\begin{itemize}
  \item For small curvature coupling, $d_H=2$, i.e. Brownian paths.
  \item For large curvature coupling, $d_H=1$, i.e. Rigid paths.
\end{itemize}
  
\figSnapShot

The simple snapshots of the paths (fig. \ref{fig:SnapShot})
visualize this change of behavior. Furthermore, the specific heat 
graph, the curvature graph and the gyration radius graph shown all 
three  a clear cross-over separating the two phases.
The large correlation in the simulation difficult a finite size 
analysis of the specific heat the cross-over to exclude the presence 
of a second order or a continuous phase transition. Nevertheless, 
it is remarkable the similarity between the specific heat graphs of 
this model and those of crystaline random surfaces, where a true 
second order phase transition has been observed. 

All our results are compatible with the existence of a critical point at
infinite curvature coupling, as it is expected from mean field analysis.

The collaboration of A. Jaramillo in the initial stages of this work is
acknowledged. Numerical computations have been performed in the 
cluster of IBM/RISC 6000 workstations of IFAE and in the CRAY of CESCA.
This work has been partially supported by research project
CICYT~AEN95/0882

\end{document}